# Nearest-Neighbor Tunneling Ansatz in the Bose-Hubbard Model

Moorad Alexanian

*Department of Physics and Physical Oceanography*
*University of North Carolina Wilmington, Wilmington, NC 28403-5606*

Email: alexanian@uncw.edu

**Abstract.** A recently introduced recurrence-relation ansatz applied to the Jaynes-Cummings-Hubbard model is here applied to the Bose-Hubbard model that reduced the model to an easily soluble model. The results obtained for the two-point density correlations resemble somewhat those obtained recently also but in a much more complicated fashion. Our ansatz may be of value for the solution of many-body quantum mechanical problems.

## 1. Introduction

In the study of many-body physics in dilute ultracold gases, the Bose-Hubbard model plays a significant role since it applies to strongly interacting gases [1] and has been experimentally realized with ultracold atoms in optical lattices [2–10]. In a recent paper [11], the many-body interference phenomena is investigated with the aid of the Bose-Hubbard model. The two-point density correlations $v$ of the number of particles in a given site shows, when varying the tunneling parameter $J/U$, a sharp peak around the value of $J/U \simeq 0.23$ where the dynamics becomes chaotic and exhibits the behavior of the indistinguishability/distinguishability of the particles [11].

Here we consider a recently introduce recurrence-relation ansatz between annihilation operators and apply it to the nearest-neighbor tunneling that considerably simplifies the Bose-Hubbard Hamiltonian [12]. This paper is structured as follows. In Sec. 2, we present the Bose-Hubbard model on an infinite, one-dimensional lattice. In Sec. 3, we introduce a recurrence-relation ansatz for the external degree of freedom associated with the nearest-neighbor of the $j$-th lattice site. In Sec. 4, we calculate the two-point density correlations of the number of particles on site $i$, irrespective of their internal states. Finally, in Sec. 5, we summarize our results.

## 2. Bose-Hubbard model

We consider the one-dimensional, infinite Bose-Hubbard model [1] with Hamiltonian

$$\hat{H} = -J \sum_{\langle i,j \rangle} \sum_{\sigma=1}^{s} \hat{a}_{i\sigma}^{\dagger} \hat{a}_{j\sigma} + \frac{U}{2} \sum_{i} \hat{N}_i (\hat{N}_i - 1). \qquad (1)$$

The first index of the creation and annihilation operators $\hat{a}_{i\sigma}^{\dagger}, \hat{a}_{j\sigma}$ refers to the Wannier orbitals of the lattice, which span the *external* single-particle Hilbert space $\mathscr{H}_{ext}$. The second index $\sigma$ refers to a basis of the $s$-dimensional *internal* single-particle Hilbert space, describing, e.g., the electronic state of an atom loaded into an optical lattice. The operator $\hat{N}_i = \sum_{\sigma=1}^{s} \hat{a}_{i\sigma}^{\dagger} \hat{a}_{i\sigma}$ counts the number of particles on lattice site $i$, irrespective of their internal state. We keep the total particle number $N = \sum_i N_i$ fixed. The two terms in $\hat{H}$ describe nearest-neighbor tunneling and on site interaction of the particles, both of which act exclusively on the external degree of freedom (d.o.f.), while the internal d.o.f. remain static.

## 3. Ansatz

Consider the following recurrence-relation ansatz for the external d.o.f. associated with the nearest-neighbor of the *j*-th lattice in (1)

$$\hat{a}_{j+1\sigma} = \hat{a}_{j\sigma} - \hat{a}_{j-1\sigma}. \tag{2}$$

and so

$$\sum_j \sum_{\sigma=1}^{s} \left[ \hat{a}_{j\sigma}^\dagger \hat{a}_{j+1\sigma} + \hat{a}_{j+1\sigma}^\dagger \hat{a}_{j\sigma} + \hat{a}_{j\sigma}^\dagger \hat{a}_{j-1\sigma} + \hat{a}_{j-1\sigma}^\dagger \hat{a}_{j\sigma} \right] = 2 \sum_j \sum_{\sigma=1}^{s} \hat{a}_{j\sigma}^\dagger \hat{a}_{j\sigma} = 2 \sum_j \hat{N}_j. \tag{3}$$

The Bose-Hubbard Hamiltonian (1) is reduced to

$$\hat{H} = -2J \sum_i \hat{N}_i + \frac{U}{2} \sum_i \hat{N}_i(\hat{N}_i - 1), \tag{4}$$

with energy eigenstates $|N_1, N_2, \cdots \rangle$ and eigenvalues

$$E = -(2J + \frac{U}{2}) \sum_i N_i + \frac{U}{2} \sum_i N_i^2, \tag{5}$$

where $N_i$ indicates the number of bosons in lattice site *i*.

We consider the simplest mixed state

$$\hat{\rho} = \lambda [a|N\rangle + b|M\rangle][\langle N|a^* + \langle M|b^*] + \beta [a|N'\rangle + b|M'\rangle][\langle N'|a^* + \langle M'|b^*] \tag{6}$$

where $|N\rangle = |N_1, N_2 \cdots \rangle$ and so $\mathrm{Tr}\hat{\rho} = \lambda + \beta = 1$ similarly for the other three eigenstates of the number operators and all four states are orthonormal. Note that $|a|^2 + |b|^2 = 1$ and so $\mathrm{Tr}\hat{\rho}^2 = \lambda + \beta = 1$ with $\gamma = \mathrm{Tr}\hat{\rho}^2 = \lambda^2 + \beta^2 \geq 1/2$, where the equality holds when $\lambda = \beta = 1/2$ and so $1/2 \leq \mathrm{Tr}\hat{\rho}^2 \leq 1$. [For S orthonormal terms in (6), instead of just two terms as in (6), one obtains $1/S \leq \mathrm{Tr}\hat{\rho}^2 \leq 1$.] In the limiting process, $\epsilon \to 0$ (see (12) below), the primed and unprimed states have the same energy and so

$$4\frac{J}{U} + 1 = \frac{\sum_i N_i^2 - \sum_i M_i^2}{\sum_i N_i - \sum_i M_i} = \frac{\sum_i N_i'^2 - \sum_i M_i'^2}{\sum_i N_i' - \sum_i M_i'} \equiv r. \tag{7}$$

Note that a finite ratio is obtained in (7) as a limiting process, viz., $\epsilon \to 0$ (see (12) below) since we are requiring $\sum_i N_i = \sum_i M_i$ in this limit.

## 4. Correlations

Consider the variance of expectation values of the two-point density correlations operator

$$\mathcal{O} = \sum_{i \neq j} (\hat{N}_i \hat{N}_j - \langle \hat{N}_i \hat{N}_j \rangle)^2 \tag{8}$$

we then have that

$$\text{Tr}(\hat{\mathscr{O}}\hat{\rho}) = |a|^2(1-|a|^2)\left[\lambda\sum_{i\neq j}(N_iN_j - M_iM_j)^2 + \beta\sum_{i\neq j}(N'_iN'_j - M'_iM'_j)^2\right]. \quad (9)$$

We now average the variable $|a|^2$ in (9) by the Dirac-$\delta$ function distribution $\delta(|a|^2 - 1/r)$ in order to require that the correlation vanishes both in the limit $J/U \to \infty$ and $J/U \to 0$ and (9) becomes

$$\widehat{\text{Tr}(\hat{\mathscr{O}}\hat{\rho})} = \frac{1}{r}\left(1-\frac{1}{r}\right)\left[\lambda\sum_{i\neq j}(N_iN_j - M_iM_j)^2 + \beta\sum_{i\neq j}(N'_iN'_j - M'_iM'_j)^2\right]. \quad (10)$$

We define

$$\nu = \frac{1}{N(N-1)}\widehat{\text{Tr}(\hat{\mathscr{O}}\hat{\rho})}. \quad (11)$$

In order to evaluate the correlation (11), we consider the following four orthonormal number states with seven lattice sites with a total number of seven bosons with at most two bosons occupying a particular lattice site,

$$\begin{aligned}|N\rangle &= |1,1,1-\frac{2r-6}{r-2}\epsilon, 2+\epsilon, 2, 0, 0\rangle, \\ |M\rangle &= |0,1,1,0,2,2,1\rangle, \\ |N'\rangle &= |1,0,1,0,2+\epsilon, 1-\frac{r-4}{r-2}\epsilon, 2, 0\rangle, \\ |M'\rangle &= |1,0,1,0,1,2,2\rangle.\end{aligned} \quad (12)$$

We then have that

$$\begin{aligned}\sum_{i\neq j}(N_iN_j - M_iM_j)^2 &= 136, \\ \sum_{i\neq j}(N'_iN'_j - M'_iM'_j)^2 &= 24.\end{aligned} \quad (13)$$

And (11) becomes

$$\nu = \frac{1}{42}\left(\frac{1}{r}-\frac{1}{r^2}\right)(136\lambda + 24\beta). \quad (14)$$

Fig. 1 shows the behavior of $\nu$ as a function of $J/U$ for different values of $\gamma = \text{Tr}\hat{\rho}^2$. Note the maximum at $J/U = 0.25$. In Ref. (11), the value of $\gamma$ is used as a measure of indistinguishability with $\gamma = 1$ for indistinguishable particles and minimal for perfectly distinguishable ones. In our case, the minimal value of $\gamma = 0.5$ owing to having only two states represented in the mixed state (6). Fig. 2 shows the same results as that in Fig. 1 albeit plotted in a linear rather than a logarithmic scale. These results are reminiscent of those in Ref. 11, where they consider the model albeit with six bosons in six lattice sites and where the six bosons can occupy a single lattice site. In our case, we limit the occupancy of a single site to only two bosons. The onset of chaos discussed in Ref. 11 would require a sort of classical limit of a quantum mechanical system that may result in the presence of a multitude of particles that would not apply to a system with a finite number of particles.

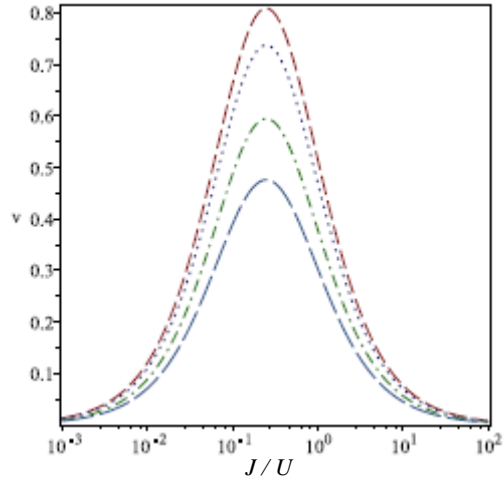

**Fig. 1.** Plot of $v$ given by (11) for different values of $\gamma = \text{Tr}\hat{\rho}^2$, $\gamma = 1.00$ (red), $\gamma = 0.81$ (cyan), $\gamma = 0.56$ (green), and $\gamma = 0.50$ (blue).

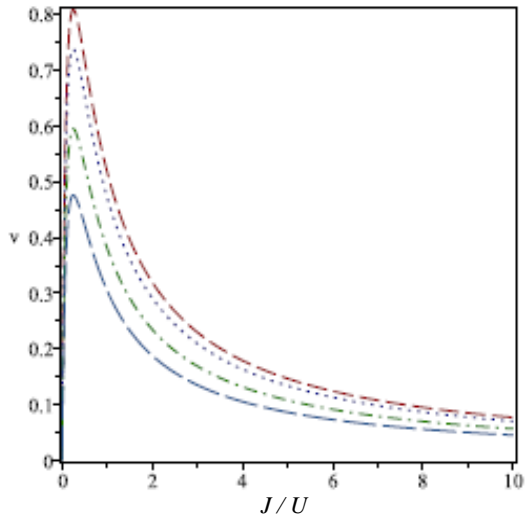

**Fig. 2**. Same plot as Fig. 1 in a linear rather than a logarithmic scale.

## 5. Conclusions

We have considered a recently introduced ansatz for the annihilation operators via a recurrence relation and applied it to the Bose-Hubbard model. The results obtained are quite consistent with previously obtained results. The simplicity of our assumption and the easily solvable resulting model indicates the potential value of this approach for solving quantum mechanical, many-body problems.